\documentstyle[epsfig,longtable]{aipproc}

\begin{document}
\title{Gamma--Ray Emission from Active Galactic Nuclei -- an Overview}\footnote{To be published in 
Proceedings of the International Symposium on High Energy Gamma-Ray  Astronomy, ed. F. Aharoninan and H. Voelk, American
Institute of Physics.}

\author{Geoffrey V. Bicknell$^*$, Stefan J. Wagner$^\dagger$ and Brent A. Groves$^*$}
\address{$^*$Research School of Astronomy and Astrophysics, Australian National University 
$\dagger$~Landessternwarte, Heidleberg}

\maketitle

\begin{abstract}
Gamma-ray emission from AGN provides us with unprecedented insights into the physics of extragalactic
jets. The emission from these jets fits naturally into the existing theoretical framework of
relativistic jets as inferred from parsec scale and kiloparsec scale observations. Models of the
$\gamma$-ray fluxes give us important knowledge of jet parameters, in particular magnetic fields and the
size of the emitting region. For example, in the case of Markarian~501, the jet plasma is subequipartiion
and the energy flux is particle dominated. Discussion of the physics of jets on the sub-parsec scale
leads naturally to questions of jet composition. The dynamics of jets are consistent with conventional
electron-proton composition if the minimum Lorentz factors,
$\gamma_{\rm min} > 100$. Jets can be electron-positron if $\gamma_{\rm min} \approx 100$.
\end{abstract}

\section*{Introduction}

One of the major discoveries of the EGRET experiment on the Compton Gamma Ray Observatory was the
detection of significant fluxes of GeV $\gamma$-rays from radio loud active galactic nuclei (AGN). Since
then ground-based TeV observations from the Whipple observatory and the High Energy Gamma-Ray Array
(HEGRA) have ushered in a new era of ground-based $\gamma$-ray astronomy at the highest energies. Apart
from one object - Centaurus A - all of the $\gamma$ rays detected so far from AGN have been from
blazars. These are highly variable AGN resulting from jets oriented close to the line of sight. The aim
of this talk is to give an overview of the physics of these objects, especially as they relate to high
energy emission and the implications for processes involving relativistic jets. A specific TeV source,
Markarian 501, is discussed in some detail. Much of this discussion leads naturally to the composition
of extragalactic jets - electron-proton or electron-positron and some recent work on this question is
also summarized here.

\section*{The Parent Objects}

In the view of AGN known as {\em Unified Schemes} the parent, or unbeamed, counterparts of blazars are
Fanaroff-Riley Class 1 and Class 2 radio galaxies (FR1 and FR2). FR1 radio galaxies contain initially
relativistic jets which decelerate to velocities of $1,000 - 10,000 \> \rm km \> s^{-1}$ on the
kiloparsec scale. Their properties on the kpc scale can be understood in terms of turbulent transonic
flows.  The deceleration from relativistic to sub-relativistic flow probably occurs on scales a few
hundred parsecs. Consistent with this, FR1 radio sources are usually asymmetric close to the core but
symmetric further out. This is interpreted as the diminishing effect of relativistic beaming. It is
widely believed, and there is good evidence for this, that BL-Lac objects are FR1 radio galaxies viewed
at small angles to the line of sight. However, sub BL-Lac objects may be beamed FR2s.

The jets in FR2 radio galaxies appear to be relativistic along their entire lengths.
Nevertheless, some deceleration is probably inevitable between the core and the hot--spots some tens of
kiloparsecs distant from the core. The appearance of FR2 jets is often dominated by knots -- interpreted
as internal shocks. They are almost always one-sided and Laing's and Garringtons's discovery
\cite{laing88a,garrington88a} that the jetted side is almost always the least depolarized, provided a
strong basis for the interpretation of their properties in terms of the beaming resulting from
relativistic motion. Again in the Unified Scheme view of the physics of these sources, FR2 radio sources
are viewed as the not highly beamed counterparts of quasars.

\section*{Components of a Theory of $\gamma$-Ray Emission from AGN}

The essential aspects of the mainstream models for production of $\gamma$-rays from AGN are the
following:
\begin{enumerate}
\item The unique association with radio--loud sources strongly indicates that the GeV--TeV emission is
the result of inverse Compton scattering of soft photons by ultra-relativistic electrons in jets moving
at relativistic bulk velocities.
\item The inferred Lorentz factors of the relativistic electrons are as high as $\sim 10^6$.
Relativistic shocks (see Achterburg; Kirk and Mastichiadis, these proceedings) are the favored mechanism
for acceleration of such high energy particles.
\item A source of soft photons is required. These can be low energy photons from the jet itself
(synchrotron self-Compton emission; SSC) or photons produced in the AGN environment (external inverse
Compton; EIC) . (It is intriguing to reflect that here, ``low energy'' can mean X-rays which had at one
stage represented the highest energy photons observed from AGN.)
\item Relativistically moving plasma boosts the observed flux density by a factor of about $10^{3-4}$.
\end{enumerate}

Apart from models of the emission itself, there are other physical processes to take into account. These
include:
\begin{enumerate}
\item The high photon density required in SSC models implies a correspondingly high opacity related to
pair production. This is especially important for TeV emission and is an important constraint on the
size of the emitting region.
\item What is the constitution of jets -- electron-ion or electron-positron?
\item What is the jet dynamics in the first $10^{16-17} \> \rm cm$? Are jets initially Poynting flux
or particle dominated?
\end{enumerate}

\subsection*{The Relative Importance of EIC and SSC}

Qualitatively, one may gain an idea of the relative importance of EIC and SSC emission from the energy
density of the soft photons. Markarian 421 and 501 are blazars in which SSC models are favored. For
those objects the radiation energy density is of order $10^{-4} - 10^{-3} \> \rm ergs \> cm^{-3}$. The
radiation energy density at a distance $10^{17}r_{17} \> \rm cm$ from an accretion disk emitting a
luminosity of $10^{44} L_{44} \> \rm ergs \> s^{-1}$ is 
\[
u_{\rm disc} \approx 0.03 \frac {L_{44}}{r_{17}^2} \> \rm ergs \> cm^{-3}
\]
This expression does not describe exactly when EIC or SCC emission may be important. For example,
\cite{sikora94a} showed that when scattered diffuse radiation from an accretion disc is taken into
account, the radiation density is enhanced by a factor $\sim \Gamma^2$.  Nevertheless, there is little
evidence in FR1s and BL-Lac objects for significant accretion disk luminosity. Hence, it is not
surprising that for BL-Lacs, SSC models are often favored. Moreover, Dr. Krawczynski has also presented
evidence at this meeting that the TeV variability of Markarian 501 is quadratic in the X-ray flux,
supporting an SSC model for that object.

\subsection*{Synchrotron Self-Compton Radiation}

Historically, synchrotron self-Compton radiation was considered in the context of the production of X-ray
radiation from radio-emitting plasma, the inverse Compton catastrophe and the limit of $10^{12} \> \rm
K$ on the brightness temperature of a synchrotron emitting plasma \cite{burbidge74}. SSC radiation at
these energies is generally produced in the Thomson limit wherein the energy of the scattered photon
$\sim \gamma^2 \epsilon$ where $\gamma$ is the electron Lorentz factor and $\epsilon$ is the energy of
the soft photon. Thus, $\gamma \sim 10^4$ electrons produce 1~keV photons from GHz photons. We know now
that higher energy photons, from infrared through to X-ray are produced from synchrotron emission at the
sub-parsec -- parsec bases of jets and the corresponding SSC photons are in the GeV to TeV range. In some
case the emission can extend beyond the Thomson limit to the Klein-Nishina regime. For the latter case
the energy of the scattered photons is of order $\gamma m_e c^2$.  Allowing for a Doppler factor of 10,
say, observed emission at $10 \> \rm TeV$ implies a Lorentz factor of about $2 \times 10^6$. It is
generally envisaged that electrons with such high Lorentz factors are produced locally at shocks. The
shocks in blazar jets are probably weak to moderate since it is not necessary to convert a large amount
of the jet energy flux into emitting particles to account for the observed fluxes.

\subsection*{Boosting of Emission from Relativistically Moving Plasma}

The Doppler factor, $\delta$ of an object moving with velocity $v = c \beta$ at an angle $\theta$ to the
line of sight, is:
\[
\delta = \frac {1}{\Gamma (1-\beta \cos \theta)}.
\]
where the bulk Lorentz factor is $\Gamma = (1-\beta^2)^{-1/2}$. Let the observed flux density from a
region of relativistically moving plasma be $F_\nu$, the luminosity distance be $D_L$ , the emissivity
at the emitted frequency, $\nu^\prime$ be $j_{\nu^\prime}$ and the element of volume in the rest frame be
$dV^\prime$, then
\[
F_\nu = \frac {\delta^3}{d_L^2} \int_{V^\prime} j_{\nu^\prime} \> dV^\prime
\]
This is one of the most fundamentally important equations of blazar physics and clearly shows that for
Doppler factors of $10-20$ that are inferred in many blazars, there is a factor of $10^{3-4}$ enhancement
in flux density over that for a Doppler factor $\sim 1$ source. This, of course, is the lynchpin of
{\em Unified Schemes} which have been so important in linking the various members of the AGN zoo.
However, it is attractive to think of this relationship in a different way and that is as a method of
using the relativistic boosting of blazars to analyze the normally faint regions close to the black hole
in radio-loud active galaxies. This is analogous to the way in which optical astronomers utilize
the amplification resulting from gravitational lensing to observe galaxies at high redshift. It is
interesting to comtemplate that both effects are consequences of the special and general theories of
relativity.

\section*{The Dynamics of Relativistic Plasma}

\subsection*{Superluminal motion}

Observations at high resolution by astronomers on physical scales ranging from parsecs to kiloparsecs
provide information that is relevant to processes occurring on the scales accessed by $\gamma$-ray
astronomy. First, of course, observations of superluminal motion have been recorded by VLBI radio
astronomers for many quasars and BL-Lac objects. Moreoever, Biretta and
colleagues have observed superluminal motion close to the nucleus of M87 using the HST
\cite{biretta99a}. The apparent superluminal velocity is of course linked to the actual velocity by the
well known expression:
\[
\beta_{\rm app} = \frac {\beta \sin \theta}{1 - \beta \cos \theta}
\]
with a maximum value of $\Gamma \beta$. Thus observations of superluminal motion provide a lower limit
on the Lorentz factor.

\subsection*{Pattern speed and jet speed}

The observation of the apparent transverse speed of a moving knot in a jet does not immediately translate
to a constraint on the velocity of the jet. The moving knots are probably shocks \cite{lind85} whose
pattern speed is likely to be different from that of the jet. For example, consider a weak jet shock
which, in the frame of the jet plasma, is moving towards the core. It is therefore swept out from the
core by the jet but at a lower velocity than the jet. For a weak shock, take $\beta_{\rm jet,sh} \approx
1/\sqrt 3$ and $\Gamma_{\rm jet,sh} \approx \sqrt{3/2}$ to be the velocity and Lorentz factor of the jet
plasma relative to the shock, $\beta_{\rm sh} \approx 1$ and $\Gamma_{\rm sh}$ the laboratory frame
shock velocity and Lorentz factor, then the transformation between the shock and laboratory frames tells
us that the lab. frame Lorentz factor of the jet is
\[
\Gamma_{\rm jet} = \Gamma_{\rm sh} \, \Gamma_{\rm jet,sh} \,
\left[ 1 + \beta_{\rm sh} \beta_{\rm jet,sh} \right] \approx 1.9 \, \Gamma_{\rm sh}
\]
The factor of $1.9$ increases for stronger shocks and underlines the fact that even weak shocks have a
substantial velocity in a relativistic plasma. 

For a jet observed almost pole-on ($\theta \approx 0$), $\delta \approx 2 \Gamma_{\rm jet} \approx 3.8
\Gamma_{\rm sh}$ for a weak shock. Hence a Doppler factor of 20, as inferred for MKN 501 (see below) is
not unexpected, given that superluminal velocities of the order of $5\,c$ are common.

\section*{Lessons from kpc scale jets}

\begin{figure}[b!]
\includegraphics[width=\textwidth]{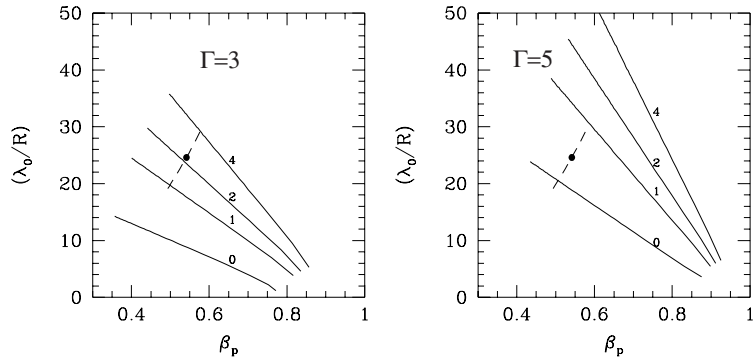}
\caption{Plot of the wavelength of the maximally growing Kelvin-Helmholtz mode against pattern speed of
the instability for jets with varying values of the parameter $\chi=\rho c^2/4p$ and values of $\Gamma =
3,5$. The point and dashed line indicates the range of values implied by the observations of the M87 jet.
}
\label{f:m87}
\end{figure}

The famous radio galaxy M87 is close enough that high resolution observations over a period of 11 years
\cite{biretta95a} have successfully detected motions of order $0.5 c$. More recent HST observations have
revealed superluminal motions close to the core. A velocity of $0.5 c$ is not fast enough to be 
consistent with the required sidedness and this motivated the interpretation of the bright knots in the
M87 jet as reverse shocks moving more slowly than the underlying jet plasma \cite{bicknell96a}. The jet
also oscillates significantly beyond knot~A. Begelman and Bicknell interpreted this as a body mode
Kelvin-Helmholtz instability. The pattern speed of the knot is the pattern speed of the instability and a
plot of maximally growing wavelength against pattern speed is shown for a $\Gamma=5$ jet in
figure~(\ref{f:m87}). The range of values indicated by the data is shown. The interesting point to note
here is that the best fit between theory and data is only attainable when
\begin{equation}
\chi = \frac{\hbox{Rest mass energy density}}{\hbox{Enthalpy}} = \frac {\rho c^2}{4 p} \sim 0 - 1
\end{equation}
This is only possible if the jet consists of electron-positron pairs or if it consists of
electron-proton plasma with a relatively high lower cutoff in the electron Lorentz factor. Specifically,
for a relativistic electron distribution described by $N(\gamma) = K_{\rm e} \, \gamma^{-a}$ for
$\gamma_{\rm min} < \gamma < \gamma_{\rm max}$, the value of $\chi$ for an electron-proton plasma is
\begin{equation}
\chi_{\rm ep} = \frac {3}{4} \frac {m_p}{m_e} \left( \frac {a-2}{a-1} \right) \, \gamma_{\rm min} 
\approx 230 \gamma_{\rm min}^{-1} \quad\hbox{for} \quad a=2.2 .
\end{equation}
Thus for M87 and electron-proton jet requires $\gamma_{\rm min} \sim 200$. On the other hand a simple
alternative is that these jets are composed of electron-positron pairs. We address this issue further
below.

\section*{Lessons from Markarian 501}

This section is the result of research carried out in collaboration with Stefan Wagner (Bicknell \&
Wagner, in preparation).

\subsection*{X-Ray and $\gamma$-ray light curves of Markarian~501}

Markarian 501 is one of the two confirmed TeV blazars, the other of course, being Markarian 421
\cite{maraschi99a}. Both of these objects are good examples of high energy SSC emission from blazar jets.
Radio images \cite{giovannini00a} show a typical one-sided core-jet structure with large apparent
jet deflection consistent with an almost pole-on jet. Two epoch observations have indicated a
superluminal velocity of $6.7 c$ for one component. However, this requires confirmation by observations
at more epochs. 

We have been undertaking analysis of multi-epoch observations of MKN~501 over a period
where it was flaring. The combined RXTE and HEGRA observations show an intriguing delay between the
1~keV X-ray minimum and the TeV minimum. (The latter occurs approximately one day later.) The reason for
this behaviour becomes apparent when one examines the evolution of the X-ray spectra above 10~keV. These
show substantial hardening at the same time as the 1~keV emission is diminishing. Clearly, these
observations caught MKN~501 at a time when one flare was subsiding and another was starting to develop.
Unfortunately, as the second flare was increasing, the moon was rising and the flare could
not be tracked. A detailed analysis of this data will soon be published. However,
some of the implications for the parameters of MKN~501 are already apparent. 

\subsection*{Constraints on parameters of emitting region}

Let us consider some of the constraints on the parameters of the emitting region which
are common to many analyses of blazars \cite{krawczynski00a,kataoka99a,maraschi99a} in the context of a
spherical geometry for the emitting region. 

\subsubsection*{Variability}

A variability time scale of
$\Delta t$ implies that the radius,
\begin{equation}
R < \frac {1}{2} \delta\,  c\,  \Delta t = 1.3 \times 10^{16} \left( \frac {\delta}{10} \right) \,
\left( \frac {\Delta t}{1 \rm day} \right)
\label{e:rvar}
\end{equation}
There are two features of this equation. First, even for relatively large Doppler factors, the size of
the emitting region is quite small, an order of magnitude smaller, for example, than the size of knots
regularly observed with VLBI. Secondly, it is large Doppler factors that save those emitting regions
which vary on a time scale of less than day, from becoming {\em too} compact and pair opaque. 

\subsubsection*{Pair opacity}

As Mattox
\cite{mattox93a} pointed out, the production of pairs by photon interactions (pair opacity) is important
for compact
$\gamma$-ray emitting regions. As Renault has also shown at this meeting (see also
\cite{krawczynski00a}), pair opacity of $\gamma$-rays on the IR background is also important. 

The threshold energy of soft photons for pair
production on $\gamma$-rays with an observed energy, $\epsilon_\gamma$, is:
\[
\epsilon_{\rm soft} =\frac {m_e^2 c^4}{\delta^{-1} \epsilon_\gamma} = 
0.26 \delta \left( \frac {\epsilon_\gamma}{\rm TeV} \right)
\]
The photon-photon cross-section peaks at the threshold energy so that a substantial photon density at
the threshold energy can be important for $\gamma$-ray absorption. Using expressions calculated by
\cite{svensson87a} for a power-law electron distribution, the pair opacity for a spherical
emitting region is:
\[
\tau \approx 2.6 \times 10^4 \left[ \frac {F_\nu(\rm keV)}{\mu \rm Jy}\right] \,
R_{16}^{-1} \, \delta^{-(2 \alpha +3)} \, z^2 \left[ \frac{\epsilon_\gamma} {\rm TeV} \right]^{0.5}
\]
It is implicit in this equation that the observed X-ray flux density is used to extrapolate the photon
density to the infrared. It is the high power of the Doppler factor in this expression which rescues TeV
blazars from potential obscurity. Note also the dependence on the radius of the emitting region,
implying a lower limit on $R$, for the detection of $\gamma$-rays. Emission at TeV energies, of course,
also provides much more severe constraints on size than emission at GeV energies.

\subsubsection*{Spectral Breaks}

Both MKN 421 and 501 exhibit spectral breaks in the X-ray. In MKN~501 the break in spectral index is
$\Delta \alpha \approx 0.3$ at about $4 \> \rm keV$. Such a break is probably the result of post-shock
cooling and this aspect of blazar physics is not strictly consistent with the assumption of a completely
homogeneous emitting region. The region following the shock must be stratified. For an homogeneous
region of finite extent, it can be shown that the superposition of cooling spectra imply $\Delta \alpha
= 0.5$. In an expanding region following a shock, the contribution of the low energies to the integrated
spectrum is diminished by adiabatic cooling and one can see, qualitatively, that the
magnitude of the spectral break is decreased. Models of this effect, could further constrain blazar
parameters. 

The location in energy of the spectral break can be estimated using the criterion:
\[
\hbox{Travel time across emitting shocked region} = \hbox{Cooling time}
\]
The travel time is estimated in the frame of the shock and one takes into account the time dilation
of plasma moving in this frame since the synchrotron cooling time, for example, is estimated from the
magnetic field in the plasma rest frame. If we take account of synchrotron cooling only, then, the
radius of the spherical emitting region is given by:
\begin{eqnarray}
R &=& \frac {3^{1.5}}{2^{3.5}} \left( \Gamma^\prime \beta^\prime \right)^{-1} \, \delta^{1/2}
\left( \frac {m_e c^2}{\sigma_T} \right) \, \left( \frac {B^2}{8 \pi} \right)^{-1} \,
\left( \frac {\epsilon}{\hbar \omega_B} \right)^{-1/2} \nonumber \\
 &=& 6.8 \times 10^{15} \, \left( \frac{\delta}{10}\right)^{1/2} \, \left( \frac {B} {0.1} \right)^{-3/2}
\, 
\left( \frac {\epsilon}{\rm keV} \right)^{-1/2} \> \rm cm
\label{e:rcool}
\end{eqnarray}
where, $(\beta^\prime,\Gamma^\prime)$ are the velocity and Lorentz factor of the plasma in the shock
frame. The quantity, $\Gamma^\prime \beta^\prime \approx 0.71$ for a weak shock and $0.35$ for a strong
shock. The numerical value is for a weak shock. Usually, the inverse Compton cooling time is comparable
to the cooling time so that the cooling length is approximately a factor of two shorter.

\subsection*{SSC fit to Markarian 501 X-Ray and TeV data}

\begin{figure}[b!]
\includegraphics[width=\textwidth]{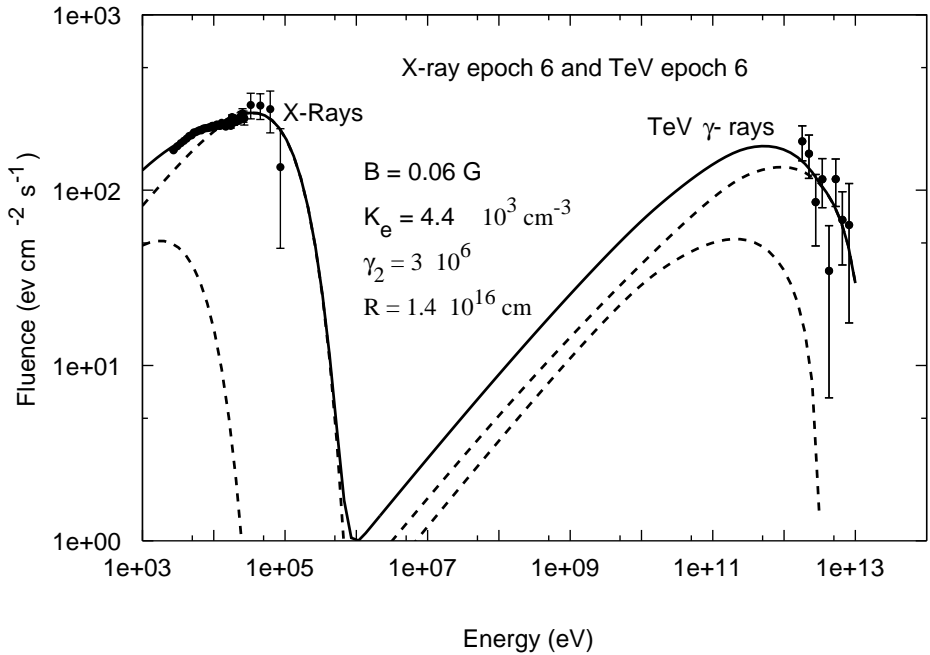}
\caption{A fit to corresponding epochs of the X-ray and TeV data for Markarian~501 for a Doppler factor
of 10. Two components are required to account for the hard and soft X-ray components.}
\label{f:MKN501}
\end{figure}

An SSC fit to  X-ray and TeV epochs of Markarian 501 at which the emission is likely to be
dominated by the rising flare is shown in figure~\ref{f:MKN501}. In this fit the TeV data have not been
corrected for intergalactic pair opacity. This increases the magnetic field and decreases the
size of the emitting region but not by a large factor.

As you can see from the fits to the MKN data two components are required to model the X-ray part of the
spectrum consistent with our earlier remarks concerning the emergence of a hard component. The cutoff
Lorentz factor, $\gamma_2$, for the hard component is determined by the peak of TeV spectrum and the fact
that the emission is in the Klein-Nishina limit. Hence the peak of the TeV spectrum is at an energy of
$\delta \gamma_2 m_e c^2$. In this fit $\gamma_2 = 3 \times 10^6$; the magnetic field is $0.06 \> \rm G$
and the radius of the emitting region is $1.3 \times 10^{16} \> \rm cm$. 

The upper limit on the radius inferred from a variability time scale of one day (see
equation~(\ref{e:rvar})) is exactly the same as the radius inferred from the model. The cooling radius
estimated from equation~(\ref{e:rcool}) above $\approx 7 \times 10^{15} \> \rm cm$. Better consistency
with the constraints imposed by cooling can be obtained with a Doppler factor
$\sim 20$  and the upper limit on $R$ implied by variability is more comfortably satisfied. However,
these parameters are probably in the right ballpark and are similar to those obtained by
\cite{kataoka99a} for an earlier flare. 

\section*{Physical Inferences from Blazar Models}

The Milano group (Celotti, Fossati, Ghisellini, Maraschi and collaborators) have also obtained fits to
numerous blazars in a similar fashion. The parameters obtained can give us valuable insight into the
physics of jets on the sub-parsec scale and again we use Markarian~501 as an example. 

\subsubsection*{Relative energy of particles and magnetic fields}

The magnetic energy density of the emitting region in Markarian 501 is approximately $1.6 \times
10^{-4} \> \rm ergs \> cm^{-3}$ and the particle energy density is approximately $3.7 \times 10^{-2} \>
\rm ergs \> cm^{-3}$. Thus the jet is subequipartition by a large factor. This has ramifications to
models of jets \cite{blandford82} in which magnetic processes drive the jets and produce a flow in which
there are equal amounts of Poynting flux and particle flux. Either issues of particle loading have to be
taken into consideration or different mechanisms for jet launching have to be considered.

\subsubsection*{Energy flux and electron-positron jets}

The energy flux of a jet consisting of relativistic particles plus cold matter is:
\[
F_E = 4 p c \beta \Gamma^2 \left[ 1 + \frac {\Gamma-1}{\Gamma} \chi \right] A
\]
where $A=\pi R^2$ is the cross-sectional area of the jet and as above $\chi = \rho c^2 / 4 p$ is the
ratio of cold matter rest energy density to enthalpy \cite{bicknell94a}. For Markarian 501,
\[
F_E \approx 7.5 \times 10^{43} \left[ 1 + \frac{\Gamma -1} {\Gamma} \chi \right] \> 
\rm ergs \> s^{-1}
\]
For $\chi \approx 0$ this is typical of (actually slightly greater than) the energy that we would
associate with a BL-Lac object or (assuming unification) an FR1 radio galaxy. Independently then, we
again arrive at the option of either an electron/positron jet or an electron/proton jet with
$\gamma_{\rm min} > 100$.  

\subsubsection*{The Gammasphere}

The notion of a gammasphere was introduced by Levinson and Blandford \cite{blandford95a} in a
theory for the EGRET sources. For the parameters relevant to Markarian 501, the pair opacity is
\[
\tau \approx 0.26 R_{16}^{-1} \left( \frac {\delta}{10} \right)^{-4} \, 
\left( \frac {\epsilon_\gamma}{\rm TeV} \right)^{0.5}
\]
The emitting region in Markarian is only just transparent for a Doppler factor of 10 and quite
transparent for a Doppler factor of 20 $(\tau \sim 0.05)$ again indicating a Doppler factor of this
magnitude. Thus the TeV $\gamma$-ray emitting region in Markarian 501 is not too far above the distance
from the black hole where it would become opaque. This raises the question, for Markarian 501, and for
other blazars, as to whether there are X-ray flares which are {\em not} accompanied by $\gamma$-rays. 

\section*{Electron-positron jets}

At two places in the above description of the physics of relativistic jets the question of jet
composition has arisen. It is therefore worth considering this question in some detail. The main issue
to be addressed here is whether an electron-positron jet can be powerful enough to supply the requisite
energy to the outer parts of a radio galaxy, quasar or BL-Lac object. An important physical constraint
is that annihilation limits the jet density and this may be inconsistent with what is required by the
energy budget \cite{ghisellini93a,celotti93a}. Celotti and Fabian cleverly worked out a way to
independently estimate the energy flux of a jet by exploiting a previously discovered correlation between
jet energy flux and narrow line region luminosity \cite{rawlings91}. Groves and Bicknell (in preparation)
have revisited this question essentially by adding extra terms to the energy flux omitted by Celotti and
Fabian.

In our sample of quasars and BL-Lac objects (mainly the sample of Celotti and Fabian with a few
additions) the inverse Compton emission from these sources is assumed to be X-ray emission and the soft
photons are from the radio part of the spectrum. (This led to some interesting discrepancies when some
well known X-ray--$\gamma$-ray blazars crept into the sample!)

The parameters of these objects were estimated following the approaches in
\cite{ghisellini93a,celotti93a}, that is:
\begin{itemize}
\item The ratio of X-ray to radio flux density is used to estimate the electorn density (specifically
the parameter $K_{\rm e}$ in $N(\gamma) = K_e \gamma^{-a}$). 
\item The synchrotron self-absorption turnover, the synchrotron flux and the VLBI angular size are
used to estimate the magnetic field and the Doppler factor.
\item  The minimum Lorentz factor was estimated from the Doppler factor.
\end{itemize}

\begin{figure}[t!]
\includegraphics[width=\textwidth]{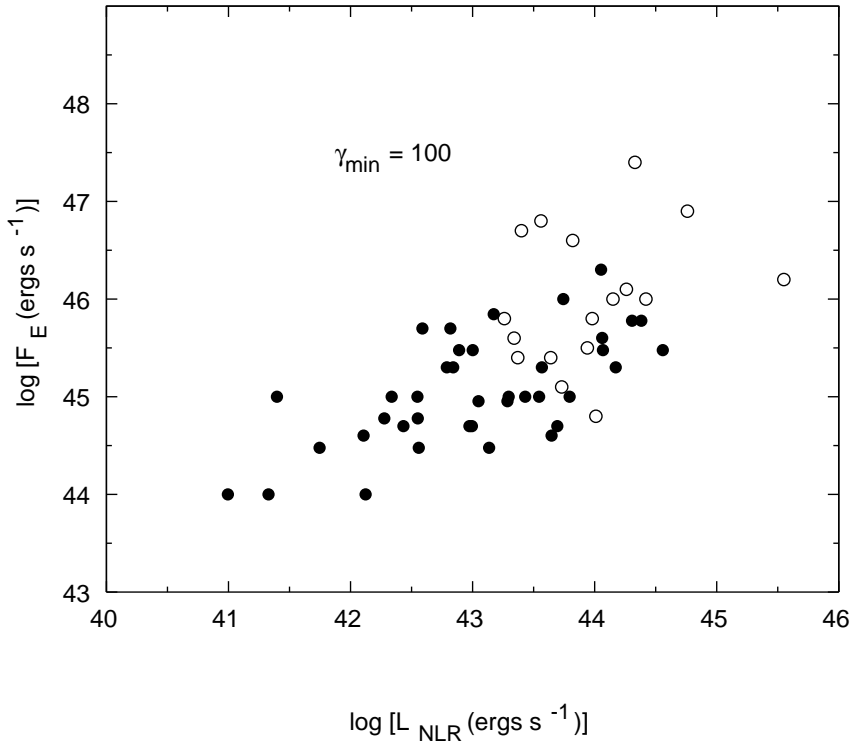}
\caption{Jet energy flux plotted against narrow line region luminosity for the Rawlings and Saunders
radio galaxies (filled circles) and for the quasars and BL-Lac objects in the expanded Celotti and Fabian
sample.}
\label{f:energy}
\end{figure}

For the present purposes we express the energy flux as:
\[
F_E = c \beta A_{\rm jet} \left[ (\Gamma^2 - \Gamma) \rho c^2 + 4 \Gamma^2 p \right]
\]
The first term represents the kinetic energy flux associated with the rest-mass density (of either cold
matter or the relativistic particles); the second term represents the enthalpy flux associated with the
relativistic particles. This term can be the most important term in the energy flux. Take $n_e$ to
be the density of both electrons and positrons then $\rho c^2 = n_e m_e c^2$ and, as before, take
$N(\gamma) = K_e \gamma^{-a}$ for $\gamma_{\rm min} < \gamma < \gamma_{\rm max}$ to describe the
distribution of Lorentz factor of the relativistic electrons and positrons. Then, the relative importance
of enthalpy and kinetic fluxes is given by
\[
\frac {4 p}{n m_e c^2} = \frac {4}{3} \frac {a-1}{a-2} \gamma_{\rm min} \sim 8 \gamma_{\rm min}
\]
for $a=2.2$. Hence, for large values of $\gamma_{\rm min}$ the enthalpy term dominates and we are able
to estimate a large energy flux for a modest density. Interestingly, the viability of an electron-positron
jet model depends on a large value of $\gamma_{\rm min}$ as does the viability of an electron-proton
model. 

Figure~\ref{f:energy} shows the plot of jet energy flux against narrow-line luminosity for
$\gamma_{\rm min} = 100$. The energy fluxes in our sample extend this correlation quite nicely and this
results from the importance of the enthalpy flux.

\begin{figure}[t!]
\includegraphics[width=\textwidth]{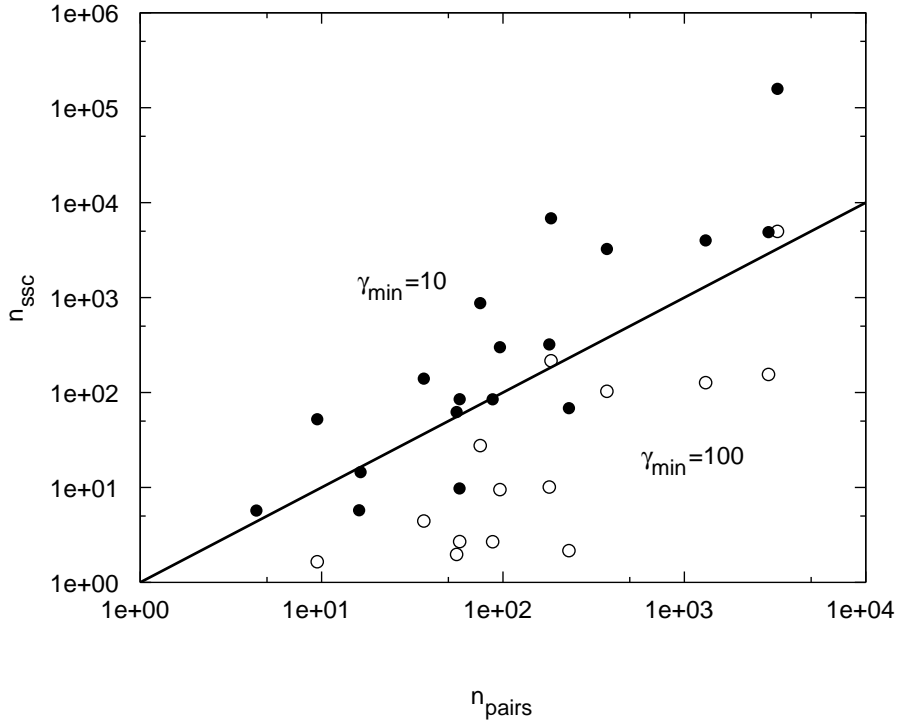}
\caption{Electron-positron density plotted against the survival density of electrons and positrons.
Filled circles correspond to $\gamma_{\rm min}=10$ and open circles correspond to $\gamma_{\rm
min}=100$.}
\label{f:npairs}
\end{figure}

The crucial plot is that of pair density. Celotti and Fabian used an estimate of the
number density of surviving pairs at the VLBI scale. Whilst this estimate is model-dependent it is
probably quite good since the main effect is the competition between annihilation and expansion. We have
used the same model and in figure~\ref{f:npairs} we plot the number density of electrons and
positrons against the survival number density, for $\gamma_{\rm min} = 10$ and 100. For
$\gamma_{\rm min}=100$ the number density is comfortably below the survival density. These two plots
therefore show that an electron-positron jet is therefore viable for this value of $\gamma_{\rm min}$.
There is a distinction here from the electron-proton case which requires $\gamma_min > 100$.

\section*{Discussion}

The gamma-ray emission from active galactic nuclei fits comfortably into the picture of relativistic jets
that has been developed from both a theoretical and observational perspective, mainly inspired
by radio observations, over the last twenty five years or so. An appealing aspect of blazars is that,
through the effects of relativistic beaming, they give us a way of investigating the physics of
relativistic jets close to the environs of black holes and lead us to confront some of the major
issues facing black hole physics in active galactic nuclei. These are: How are jets produced; what
are the physics of jet propagation in the first $10^{16} \> \rm cm$ and why do some AGN produce jets and
why do some not? The developing field of $\gamma$-ray astronomy, one of the last frontiers, is well
placed to provide important information on these issues. One looks forward to the next generation of
$\gamma$-ray telescopes and to attempting to resolve many of the problems with which they will
surely provide us.

GVB is grateful to the organisers of this meeting for assistance in funding his visit to Heidelberg.


\begin{thebibliography}{10}

\bibitem{laing88a}
Laing R.~A.,
{\em Nature}, {\bf 331}, 149 (1988).

\bibitem{garrington88a}
Garrington S.T,  Leahy J.P., Conway R.G., and Laing  R.A.,
{\em Nature}, {\bf 331}, 147 (1988).

\bibitem{sikora94a}
Sikora M., Begelman M.C., and Rees M.J., {\em ApJ}, {\bf 421}, 153 (1994).

\bibitem{burbidge74}
 Burbidge G.R.,  Jones T.W., and O'Dell S.L.,
{\em ApJ}, {\bf 193}, 43, (1974)

\bibitem{biretta99a}
Biretta J.A., Sparks W.B., and Macchetto F.,
{\em ApJ}, {\bf 520}, 621 (1999)

\bibitem{lind85}
 Lind K.R., and Blandford R.D.,
{\em ApJ}, {\bf 295}, 358 (1985)

\bibitem{biretta95a}
 Biretta J.A., Zhou F., and  Owen F.N.,
{\em ApJ}, {\bf 447}, 582 (1995)

\bibitem{bicknell96a}
Bicknell G.V., and Begelman M.C.,
{\em ApJ}, {\bf 467}, 597 (1996).

\bibitem{maraschi99a}
Maraschi L. et al., 
{\em ApJ}, {\bf 526}, 81 (1999).

\bibitem{giovannini00a}
Giovannini G., Cotton W.D., Feretti L., Lara L., and Venturi T.,
{\em Advances in Space Research}, {\bf 26}, 693 (2000)

\bibitem{krawczynski00a}
Krawczynski H.,  Coppi P.S., Maccarone T., and  Aharonian F.A.,
{\em A\&A}, {\bf 353}, 97 (2000).

\bibitem{kataoka99a}
Kataoka J. et al.
{\em ApJ}, {\bf 514}, 138 (1999)

\bibitem{mattox93a}
Mattox J.R.  et al.
{\em ApJ}, {\bf 410}, 609 (1993)

\bibitem{svensson87a}
Svensson R.,
{\em MNRAS}, {\bf 227}, 403 (1987)

\bibitem{blandford82}
 Blandford R.D., and  Payne D.~G.,
{\em MNRAS}, {\bf 199}, 883 (1982)

\bibitem{bicknell94a}
Bicknell G.V.,
{\em ApJ}, {\bf 422}, 542 (1994)

\bibitem{blandford95a}
Blandford R.D., and Levinson A.,
{\em ApJ}, {\bf 441}, 79 (1995).

\bibitem{ghisellini93a}
Ghisellini G., Padovani P., Celotti A., and Maraschi L.,
{\em ApJ}, {\bf 407}, 65 (1993)

\bibitem{celotti93a}
Celotti A., and  Fabian A.C.,
{\em MNRAS}, {\bf 264}, 228 (1993)

\bibitem{rawlings91}
Rawlings S.R., and Saunders R.,
{\em Nature}, {\bf 349}, 138 (1991)

\end{thebibliography}
\end{document}